\documentclass[11pt,a4paper]{article}
\pdfoutput=1
\usepackage{jheppub}
\usepackage{extarrows}
\usepackage{caption}
\allowdisplaybreaks
\usepackage{subfigure}
\usepackage{slashed}
\usepackage{cancel}
\usepackage[normalem]{ulem}
 \usepackage{setspace} 
 \usepackage{amsmath}
\usepackage{caption}
\usepackage{xcolor}  
\usepackage{color}
\allowdisplaybreaks
\def\bea{\begin{eqnarray}}
\def\eea{\end{eqnarray}}
\def\be{\begin{equation}}
\def\ee{\end{equation}}

\usepackage[utf8]{inputenc}

\DeclareUnicodeCharacter{2212}{-}
\usepackage{comment}
\usepackage{multirow}
\usepackage{makecell}

\usepackage[T1]{fontenc}

\newcommand{\beq}{\begin{equation}}
\newcommand{\eeq}{\end{equation}}
\newcommand{\beqa}{\begin{eqnarray}}
\newcommand{\eeqa}{\end{eqnarray}}

\usepackage{cancel}
\usepackage{soul}

\vspace*{-2cm}

\title{Testing the lepton content of the proton at HERA and EIC\footnote{\large Dedicated to the memory of James D. Bjorken.}}
  \author[a]{Leandro Da Rold,}
  \author[b]{Anibal D. Medina,}
  \author[c]{Subhojit Roy,}
  \author[c,d,e]{Carlos E.M. Wagner}
\affiliation[a]{Centro At\'omico Bariloche, Instituto Balseiro and CONICET, Av. Bustillo 9500, 8400, S.C. de Bariloche, Argentina}
\affiliation[b]{IFLP, CONICET - Dpto. de F\'isica, Universidad Nacional de La Plata, C.C. 67, 1900 La Plata, Argentina}
\affiliation[c]{HEP Division, Argonne National Laboratory, 9700 Cass Ave., Argonne, IL 60439, USA}
\affiliation[d]{Enrico Fermi Institute, Physics Department, University of Chicago, Chicago, IL 60637, USA}
\affiliation[e]{Kavli Institute for Cosmological Physics, University of Chicago, Chicago, IL 60637, USA}

    \emailAdd{daroldl@ib.edu.ar} \emailAdd{anibal.medina@fisica.unlp.edu.ar}
    \emailAdd{sroy@anl.gov}
    \emailAdd{cwagner@uchicago.edu}
    \preprint{EFI 24-8} 
  \abstract{Although protons are baryons with an overall vanishing lepton number, they possess a non-trivial leptonic content arising from quantum fluctuations which can be described by lepton parton
distribution functions (PDFs) of the proton.
These PDFs have been recently computed and can be used to define lepton-induced processes at high-energy colliders.
In this article, we propose a novel way to test the computation of lepton PDFs of the proton by analyzing both non-resonant di-lepton and resonant 
Z gauge boson production processes induced by leptons within the proton at proton-electron colliders like HERA and EIC.
Despite the fact that lepton PDFs of the proton are known to be small, this work demonstrates that both processes imply a measurable yield of events at HERA and EIC, which could be used to test these PDFs. 
}


\begin{document}
\maketitle
\flushbottom 
\section{Introduction}
Parton Distribution Functions (PDFs) \cite{Bjorken:1968dy,Feynman:1969wa,Bjorken:1969ja,Altarelli:1977zs} are important for hadron collider phenomenology as they describe the momentum distribution of the constituents within the proton. These PDFs are essential in the computation of the cross-sections for processes mediated by strong interactions when the hard ultraviolet (UV) dominated scattering factorizes from the infrared (IR) behaviour that is encapsulated in the proton's PDFs. Due to their non-perturbative nature, the proton's PDFs are obtained from the analysis of experimental data.
Quantum fluctuations inside the proton imply that not only valence quarks contribute to the proton's PDFs but also sea quarks and gluons via Quantum Chromodynamic (QCD) interactions, and the photon and ultimately leptons via Quantum Electrodynamic (QED) interactions of the Standard Model (SM).

It has also been argued that an analogue concept of PDFs can be applied to colliding charged leptons, as in the case of $pe$ collision.
Even though electrically charged leptons are elementary particles in the SM, they can undergo  collinear initial state radiation (ISR), characterized by  transverse momentum ($P_T$) significantly smaller than the energy of the hard scattering process in $pe$ collisions.
As the energy of the lepton beam increases, the ISR becomes significant, not only reducing the colliding energies of the leptons but also initiating new interactions of the radiation fields.
At leading order (LO) the emission of collinear photons by high-energy charged leptons can be described using the equivalent photon approximation (EPA),
given by the Weizs\"acker-Williams spectrum~\cite{vonWeizsacker:1934nji,Williams:1934ad} 
\begin{equation}
f_{\gamma,l}(x)  \approx  Q_{\ell}^2 \frac{\alpha}{ 2\pi} P_{\gamma,\ell}(x) \ln{E^2\over m^2_{\ell}}, 
\label{eq:EPA}
\end{equation}
where $E$, $m_{\ell}$ and $Q_{\ell}$ are the energy, mass  and charge of the lepton, respectively and the splitting function,
\begin{equation}
P_{\gamma,\ell}(x) = \frac{1+(1-x)^2}{x} \,, 
\label{splteq1}
\end{equation}
describes the 
probability of the lepton to emit a photon ($\ell \to \ell \gamma$) with a fraction $x$ of its longitudinal momentum.
 Subsequently, those radiated photons can furthermore provide other fermion parton content to the colliding lepton via  the $\gamma \to  f \bar{f}$ splitting function,
 \begin{equation}
  P_{f, \gamma}(z_{f}) = 1-2z_{f} +2z_{f}^2 \, 
\label{splteq2}
\end{equation}
where $z_{f}$ is the fraction of momentum of the fermion with respect to the photon.
The EPA description breaks down at high energies, when $E \gg m_{\ell}$, implying that the collinear logarithm $(\alpha/2\pi) \ln{(E^2 / m^2_{\ell})}$ must be resummed and a factorization scale is introduced.
Ensuring that the physics remains invariant under changes in the factorization scale leads to the QED analogue of the Dokshitzer-Gribov-Lipatov-Altarelli-Parisi (DGLAP) Eqs.~\cite{Gribov:1972ri,Dokshitzer:1977sg,Altarelli:1977zs} and to the generalization of PDFs for charged leptons, recently computed in Refs.~\cite{Garosi:2023bvq,Han:2020uid,Han:2021kes,Greco:2016izi}.

A similar description can be applied to the QED interactions within the proton~\cite{Spiesberger:1994dm,Roth:2004ti,Martin:2004dh}.
However, in this context, when estimating the proton PDFs involving QED-induced processes, hadronization effects lead to a modification of the logarithmic factor in Eq.~(\ref{eq:EPA}) to  $\ln(\mu_F^2/\Lambda^2)$, where $\mu_F$ denotes to the factorization scale and $\Lambda$ represents the typical hadronic scale~\cite{Buonocore:2020nai}.
These ideas were put into practice in Refs.~\cite{Manohar:2016nzj,Manohar:2017eqh} in what is known as the LUX approach, where it was shown that the photon PDF of the proton could be obtained in a model-independent way using the structure functions and form factors derived from the deep-inelastic scattering data of proton-electron ($pe$) colliders, such as the Hadron-Electron Ring Accelerator (HERA) collider.
These computations resulted in a much more precise determination of the photon PDF, reducing the uncertainty of previous analyses that were either model-driven or solely data-driven~\cite{Martin:2004dh, Ball:2013hta}. Such an approach has now been integrated into various global PDF sets~\cite{Bertone:2017bme, Cridge:2021pxm,Xie:2021equ,Xie:2023qbn,NNPDF:2024djq}.
Recently, this has been extended to the calculation of the lepton PDFs of the proton up to next-to-leading order (NLO)~\cite{Buonocore:2020nai}.

Since lepton PDFs of the proton are significantly smaller than the colored parton ones, the lepton-initiated processes at hadron colliders tend to be associated with small cross-sections and suffer from large backgrounds.
Thus in this article, we propose to test the estimation of the lepton PDFs of the proton~\cite{Buonocore:2020nai} by studying non-resonant di-lepton and resonant $Z$-gauge boson production at the HERA and the futuristic Electron-Ion Collider (EIC).  Such processes have been measured by the HERA experiments, obtaining good agreement with the SM predictions in the kinematical regions that they explored~\cite{ZEUS:2008goe,ZEUS:2003vfj,Kerger:2000eh,H1:2006wzo,ZEUS:1999qvb,ZEUS:2003nmf,Diener:2002if,ZEUS:1997bxs,ZEUS:2002pdo,ZEUS:2008hcd,H1:2009rfg,H1:2009mij,Nobe:2014csa}. We will show how to extract information on the lepton PDFs of the proton from these measurements, concentrating on particular regions of parameter space that can be probed with the HERA stored data and from future EIC data.
We study the lepton-initiated non-resonant di-lepton production processes, which involve the lepton PDFs of the proton ($p (\ell) e^{-}\rightarrow \ell \ell$)~\footnote{In this work, we use the convention $i(j)$ to indicate that the process involves `$j$'-th parton of `$i$'. For example, 
$p(l)$ denotes the lepton parton of the proton. However, for the electron parton of the electron, we simply refer to it as $e^{-}$, instead of using the convention $e^{-} (e^{-})$.}, where $\ell = e, \mu$, as well as the photon-initiated processes ($p (\gamma) e^{-}(\gamma) \rightarrow \ell \ell$), which involve the photon PDFs of the electron and the proton.
In addition to that, we also study the resonant $Z$-boson production through the lepton-initiated process ($p (e^{\pm}) e^{\mp} \rightarrow Z$) that involves the electron  PDFs of the proton and the quark-initiated processes ($p (q) e^{-} (q) \rightarrow Z$) which involve the quark PDFs of proton and electron. Notice that lepton-initiated $Z$-resonant production process ($p(\ell^{\pm}) p(\ell^{\mp})\rightarrow Z$) cannot be tested at hadron colliders, as the quark-initiated processes ($p (q) p(\bar q) \rightarrow Z$) dominate, making the $pe$ collider a unique machine to study lepton-initiated $Z$-resonant production and test the lepton PDFs of the proton.
 
\begin{figure}[t!]
\begin{center}
\includegraphics[width=0.74\linewidth]{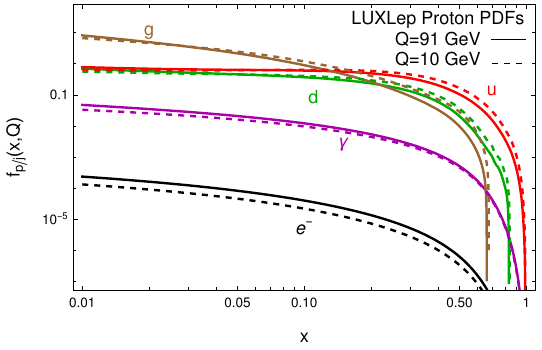}

\includegraphics[width=0.74\linewidth]{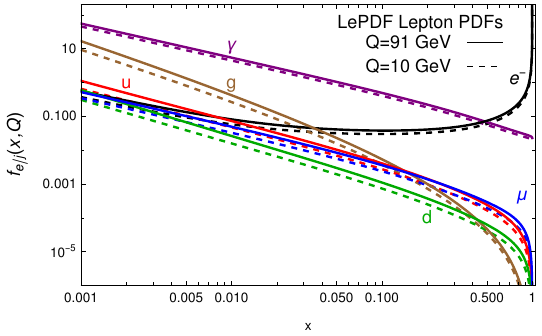}
\caption{PDFs of the proton (top) and electron (bottom) evaluated at the renormalization scale $Q = 91$~GeV (continuous) and $10$~GeV (dashed), using LUXLep set~\cite{Buonocore:2020nai} and LePDF set~\cite{Garosi:2023bvq}, respectively. Various colors denote distributions of up quark (red), down quark (green), photon (purple), gluon (brown) and electron (black).} 
\label{proton-elec-pdf}
\end{center}
\end{figure}

This paper is organized as follows: In section~\ref{proton_electronpdf} we discuss the PDFs of proton and electron.
A discussion on the NLO corrections of various processes, that we studied in this work, is provided in section~\ref{NLO}. 
The results in the non-resonant di-lepton production and the resonant electroweak gauge boson production at HERA are discussed in sections~\ref{non-resonant} and \ref{resonant_production}, respectively. The results for these processes at the EIC are discussed in section~\ref{EIC}.
The final section~\ref{Conclusions} provides the conclusions of this work.

\section{Proton and Electron PDFs}
\label{proton_electronpdf}
The PDFs of the proton as a function of $x_{p/i}$, the longitudinal momentum fraction of the constituent parton $i$, from the LUXLep set~\cite{Buonocore:2020nai} are shown in the top plot of Fig.~\ref{proton-elec-pdf}, using ManeParse~\cite{Clark:2016jgm}. We set the renormalization scale $Q$ equal to the factorization scale and choose  $Q=10$ and 91 GeV, of the order of the expected relevant scales for non-resonant di-lepton and resonant $Z$-boson production, in dashed and solid lines, respectively.  
While the PDFs of the quarks are $\mathcal{O}(1)$ over a significant portion of the low $x_{p/q}$ range, those for photons and leptons are suppressed by approximately $\mathcal{O}(10^{-1})$ and $\mathcal{O}(10^{-3})$, respectively, relative to those quark PDFs.
By analogy, the bottom plot of Fig.~\ref{proton-elec-pdf} illustrates the electron PDFs evaluated at the same renormalization scales from the LePDF set~\cite{Garosi:2023bvq}, as a function of $x_{e/i}$. Notably, this plot reveals that even in the region near $x_{e/i} \simeq 1$, where the electron carries most of the momentum, the photon PDF of the electron can still provide significant contributions.

The order of magnitude of the cross-sections for the various processes of interest can be estimated using power counting arguments similar to those employed in Ref.~\cite{Buonocore:2020nai} for the PDFs.
We consider factorization for calculating the cross-sections of various processes in $pe$ collisions, expressed as $f_{p/i} \otimes f_{e/j} \otimes \hat\sigma_{ji}$, where $\hat\sigma_{ji}$ represents the hard scattering between partons $i$ from the proton and $j$ from the electron, with $f_{p/i}$ and $f_{e/j}$ being the respective parton PDFs.

The power counting argument is broadly characterized by two parameters: $\alpha_s$ and $L$, where $L=\log\mu_f^2/\Lambda^2$, as argued below Eq.~(\ref{eq:EPA}). Moreover, one can consider the approximations $\alpha\sim\alpha_s^2$ and $L\sim\alpha_s^{-1}$ for the processes described below. 
The quark and gluon PDFs in the proton, $f_{p/q}$ and $f_{p/g}$, are of order one, since at the perturbative level they are given by a sum of terms of order $(\alpha_sL)^n$, all of  ${\cal O}(1)$. On the other hand, the photon and lepton PDFs of the proton are: $f_{p/\gamma}\sim \alpha_s$ and $f_{p/\ell}\sim\alpha_s^2 C_\ell$, where $C_\ell$ is a constant factor depending on the charges and flavors of the partons, $C_\ell=3/20$ for our analysis~\footnote{To the LO, the photon splitting function $P_{f,\gamma}$, Eq. (\ref{splteq2}), is weighted by the charge and multiplicity of the fermion anti-fermion pair, giving a suppression factor for a charged lepton:$C_\ell=Q_e^2/(3Q_e^2+2N_cQ_u^2+3N_cQ_d^2)$~\cite{Han:2020uid}.}. 
Applying these estimates to the resonant $Z$ production cross-section from $e^+e^-$ collision, with $\hat\sigma_{e^+e^-}\sim\alpha/\hat s\sim\alpha_s^2/\hat s$, where $\hat s$ is the partonic centre of mass energy of order $m_Z^2$, leads to an estimate for the cross-section for the lepton-initiated process of order $\alpha_s^4C_\ell/m_Z^2$.
Regarding the  quark-initiated processes while the electron PDF $f_{e/e}$  is of $\mathcal{O}(1)$, the photon and anti/quark PDFs $f_{e/\gamma}$ and $f_{e/q}$ can be estimated to be of order $\alpha  L'\sim \alpha_s$ and $(\alpha  L')^2C_q\sim \alpha_s^2 C_q$, respectively, with $L'=\log(\mu_f^2/m_e^2)$ and $C_q\sim 3 Q_q^2/20$.
Since resonant $Z$ production via quark-antiquark annihilation ($q\bar{q} \to Z$) is of order $\alpha$, the process initiated by the $\bar{u}$ and $\bar{d}$ contents of the electron is of order $\alpha_s^4 (C_u + C_d) / m_Z^2$, leading to a competitive cross-section compared to lepton-initiated processes. Similar estimates for non-resonant di-lepton production lead to cross-sections of order $\alpha_s^6/\hat t$ with $\hat t\sim P_T^2$, both for lepton- and photon-initiated processes. With characteristic $P_T \gtrsim \mathcal{O}(10)$ GeV from the selection cuts employed at HERA, it leads to a production rate similar to the resonant $Z$-boson production process. 

There is a relevant quantity that is  very useful in distinguishing lepton-initiated processes in the production of di-leptons and Z gauge bosons from all other SM contributions which we treat as background:
$E-P_z\equiv \sum_i (E^{i} - P_{z}^{i})$, where the sum runs over all detected particles. Here, $E^i$ ($P_{z}^{i}$) represents the energy (momentum along the proton beam axis) of the '$i$'-th final state particle. According to longitudinal momentum conservation, this quantity should be equal to $2 x_{e/j} E_e$,
where $j$ represents the colliding parton within the electron. Hence, a cut of $E - P_z$ 
close to 2$E_e$ leads to enhancing the purely electron-initiated process
due to the almost Dirac delta-like behaviour of the electron PDFs of electron near $x_{e/e} \sim 1$.
On the other hand, the kinematical distributions of the quark-initiated  and photon-initiated processes predominantly peak at relatively small $x_{e^{-}/q}$ and $x_{e^{-}/\gamma}$, which corresponds to significantly smaller $E-P_z$ values,  implying that a cut of $E - P_z$ 
close to 2$E_e$  also reduces 
significantly their contribution.

%
\section{NLO corrections}
\label{NLO}

There are virtual corrections at the one-loop level and real corrections from radiation to the processes that we consider.
For resonant $Z$ production,  NLO QED corrections can be obtained via the photon initiated processes: $\gamma e^-\to Ze^-$, with a partonic cross-section: $\hat\sigma_{\gamma e^-}\sim\alpha^2/m_Z^2\sim\alpha_s^4/m_Z^2$, that after convolution with the PDFs leads to a cross-section of order $\alpha_s^5 C_{\ell}/m_Z^2$ and hence smaller than the ones studied in this article.  Corrections from processes with virtual particles in a loop, as well as those with a final photon: $e^+e^-\to Z\gamma$, are of order $\alpha_s^6 C_{\ell}/m_Z^2$, and hence suppressed. The NLO QED correction: $q\gamma\to q Z$, after convolution with $f_{e/\gamma}$ is of order $\alpha_s^5/m_Z^2$, and contains extra hard jet activity in the forward direction. 
On the other hand, NLO QCD corrections in $Z+j$ production: $gq\to Zq$ and $q\bar q\to Zg$, where the initial quark or gluon coming from the colliding lepton, are of similar order.
However, since for the lepton-initiated resonant process $x_{e^-/e^-}\simeq 1$, whereas for the former processes the kinematical distributions peak at low values of $x_{e^-/q,g}$, the latter can be controlled by a $E-P_z$ kinematic cut. Finally, processes with a final photon: $q\bar q\to Z\gamma$ are suppressed by $\alpha_s^2$ compared with the resonant production. A similar estimate can be done for the production of $W^{\pm}$ gauge boson. 

 For the non-resonant di-lepton production there are real corrections from photon emission of initial and final particles, with the same initial state as the leading processes, that are suppressed by $\alpha_s^2$ compared with them. Di-lepton plus jet final states can be obtained from $\gamma q\to \ell\ell q$, being suppressed by $\alpha_s$ only, compared with the leading processes.
 When considering the quark and gluon content of the electron one can also obtain di-lepton plus jet final states which are suppressed by $\alpha_s$. The kinematic distributions of processes initiated by partons of the electron, other than the electron itself, are peaked towards low values of $E-P_z$, and thus the rates of these processes can be controlled by cuts on this variable. Virtual QED corrections are expected to be suppressed by $\alpha_s^2$ compared with the real ones.

A dedicated analysis containing these corrections, as well as parton showering, is needed for a precise calculation.  
Dependence of the cross-section on the renormalization and the factorization scales is expected to decrease after including higher-order corrections. Therefore, an estimate of these corrections can be obtained by examining the scale variation of the cross-section. As we will see later, our LO calculation of the cross-sections varies by a few tens of percent when the scale is varied by a factor of 2 up and down relative to a reference scale. This variation is not surprising, given that we are performing an LO calculation, for which large scale variations are expected. Although the PDFs are evaluated at NLO, the hard scattering processes are calculated at LO. The scale variation of the PDFs, which is included in our cross-section evaluations, is estimated to be less than 4\% for the $x$ values of interest~\cite{Buonocore:2021bsf}. The optimal choice for the reference scale can only be determined once the actual NLO calculation is performed.
To select the reference scale, we will take guidance from Ref.~\cite{Buonocore:2021bsf}, which showed that at the LHC, the LO evaluation agreed more closely with the NLO result when the scale was chosen to be a factor of 2 smaller than the natural scale of the process. The natural scale is defined as the scalar sum of the transverse momenta of the final states for non-resonant production, and the mass of the resonantly produced particle for resonant production processes. These processes are discussed in detail in the following sections.

Before discussing these processes, it is important to point out that the current PDF implementation in Luxlep does not support proper showering calculations with {\tt PYTHIA8}~\cite{Sjostrand:2014zea}, since it cannot handle incoming leptonic partons. 
Recently there have been a few examples using the POWHEG method at NLO with parton showering at the LHC~\cite{Buonocore:2021bsf,Buonocore:2022msy,Almeida:2022udp, Korajac:2023xtv}, but up to our knowledge it has not been implemented for $pe$ colliders.
Since all the signal processes we compute are mostly weakly interacting, we do not expect showering to introduce significant corrections to these processes. 

\section{Non-resonant di-lepton production}
\label{non-resonant}
We consider initially non-resonant di-lepton production at HERA. The signal is dominated by photon-mediated $t/u$ and $s$-channel processes and we concentrate on the $e e$, $e \mu$ and $\mu \mu$ final states.  As stressed previously, the purely electron-initiated process from the colliding electron becomes more relevant in the region of  $x_{e/i} \simeq 1$, thus in part of our analysis we will consider values of $E - P_z$ close to 2$E_e$.
Due to the nature of the lepton PDF of the proton, we expect di-lepton final states with both opposite and same charges, as well as different flavors.
For same-flavor, opposite-sign final states ($e^{+}e^{-}$ and $\mu^{+}\mu^{-}$), significant contributions also come from the photon-initiated processes.
The lepton flavor violating final state ($e\mu$) and the same-sign di-lepton final states are much cleaner since photo-production processes do not contribute to these channels. An intriguing aspect of the $e\mu$ signal is that it exclusively involves the muon PDF of the proton, allowing for independent probing without the assumption of flavor universality. 

HERA originally operated with an $e^{\pm}$ beam of 27.6 GeV and a proton beam of 920 GeV with 0.496~$\text{fb}^{-1}$ of integrated luminosity. A dedicated analysis considering events with at least two central (20$^\circ$ $< \theta <$ 150$^\circ$) leptons (electrons or muons) was already performed using the combined data from the H1 and ZEUS detectors at HERA with a total integrated luminosity of 0.94~fb$^{-1}$~\cite{H1:2009mij}.
In the selected events one of the leptons must have $P_T^{\ell} >$ 10 GeV and the other $P_T^{\ell} >$ 5 GeV and all lepton candidates are isolated with $\Delta R > 0.5$. An
$E-P_Z < 45$~GeV cut is applied to isolate the di-leptonic ($ee$ and $\mu \mu$) photo-production regime, in which the virtuality $Q^2$ of the photon, emitted by the colliding electron, is low and the scattered beam $e$ is lost in the beampipe. We perform a simplified partonic level analysis using the PDF formalism for the colliding $e$ and $p$ at LO and compare our results with the observed data.
Under these kinematical cuts, the non-resonant contribution to the di-lepton final states primarily originates from $0.005\lesssim x_{p/e,\mu}\lesssim 0.03$ region of the lepton PDFs of the proton (both $e$ and $\mu$ flavors).
In addition to these cuts, in our analysis we also consider the detector efficiencies~\cite{H1:2009mij} for $e$ and $\mu$ in the central region: 80\% and 90\% at HI and ZEUS for $e$ and 90\% and 55\% at HI and ZEUS for $\mu$, respectively. We assume for simplicity $pe^-$ collisions only since it is straightforward to obtain the results for $pe^+$ collisions.

In our study, we compute all cross-sections at LO using~{\tt MadGraph5-aMC$@$NLO-v3.5.1}~\cite{Alwall:2014hca}.
We use the lepton, quark, and gluon PDFs of the proton from the Luxlep-NNPDF31$\_$nlo$\_$as $\_$0118$\_$luxqed PDF~\cite{Buonocore:2020nai}, and the quark and photon PDFs of the electron from the LePDF set~\cite{Garosi:2023bvq}.
To discuss the general behavior of signal and background, we work at parton-level, expecting that our conclusions will still hold when proper showering and hadronization are taken into account, and we analyze events using Madanalysis5~\cite{Araz:2020lnp}.

\begin{table}[t!]
\centering
\begin{tabular}{|c|c|c|c|}
\hline\rule{0mm}{5mm}
final state & $\ell$-init & $\gamma$-init & $\gamma$-init$_{E-P_z>45{\rm GeV}}$ \\ 
\hline\rule{0mm}{7mm}
$e^+e^-$ & $175^{+41}_{-34}$ & $505^{+55}_{-53}$ & $62^{+8}_{-7}$ \\[0.2cm]
\hline \rule{0mm}{7mm}
$e^-e^-$ & $282^{+64}_{-49}$ & $-$ & $-$ \\[0.2cm]
\hline \rule{0mm}{7mm}
$\mu^+\mu^-$ & $10^{+2}_{-2}$ & $387^{+43}_{-40}$ & $48^{+6}_{-6}$\\[0.2cm]
\hline \rule{0mm}{7mm}
$e^-\mu^- + e^-\mu^+$ & $ 284^{+64}_{-53}$ & $-$ & $-$ \\[0.2cm]
\hline
\end{tabular}
\caption{Estimated number of di-lepton events at HERA with an integrated luminosity of 0.94~fb$^{-1}$~\cite{H1:2009mij}, using MG5. This is after applying all the phase space cuts that are considered in Ref~\cite{H1:2009mij}.
The uncertainty is determined by adjusting the factorization scale, which is set to the $P_T$ of the lepton, up and down by a factor of 2.}
\label{table-MG5}
\end{table}
We can compare our predictions for the number of events originating from lepton-initiated and photon-initiated processes in the different di-lepton channels with the existing analysis carried out using the HERA data~\cite{H1:2009mij}. 
Our results are summarized in Table~\ref{table-MG5}.
As discussed in section~\ref{NLO}, we set the factorization scale $Q$ to half the scalar $P_T$ sum of the final state leptons and estimate the cross-section uncertainties by varying $Q$ by a factor of 2 up and down.

Within the error margins, we find agreement with the observed data for the production of $ee$ and $\mu\mu$ events via lepton-initiated and photon-initiated processes in the region of $E - P_z > 45$ GeV~\footnote{The lepton-initiated processes correspond to the exchange of $\gamma$ in the $s$ and $t$ channels for $ee$ and in the $s$ channel for $\mu \mu$, whereas photon-initiated processes involve the exchange of leptons in the $t$ channel.}, where the lepton-initiated signal peaks and the photo-production contribution remains comparatively small. Our results, derived by summing the values in the first and third rows of Table~\ref{table-MG5}, predict approximately $519^{+113}_{-90}$ events for the $ee$ final state and $58^{+8}_{-8}$ for the $\mu\mu$ final state. These predictions compare with the experimentally reported counts of 418 and 63 events, respectively, corresponding to deviations of $0.9\sigma$ for $ee$ and $0.3\sigma$ for $\mu\mu$. The deviations are calculated using Poisson errors ($\sqrt N$) as an estimate for experimental uncertainty, given that explicit errors were not provided by the experiments.

The agreement is less favorable  in the flavor-violating $e\mu$ final state, which is anticipated to be the cleanest channel, with our estimate of $284^{+64}_{-53}$ events compared to the 173 events observed, corresponding to a $1.7 \sigma$ deviation. 
Similar over estimations are observed in di-lepton pair production via the photon-initiated process in the $E - P_z < 45$~GeV region, as shown by subtracting the numbers in the second and third columns of Table~\ref{table-MG5}. The rate of these processes involves the photon PDFs of the electron and the proton. We estimate $443^{+47}_{-46}$ events for $ee$ and $339^{+37}_{-34}$ for $\mu\mu$ via the $(\gamma\gamma)_{E-P_z < 45 \text{GeV}}$ process, compared to the experimentally reported 284 and 235 events, respectively. This corresponds to deviations of $2.5\sigma$ for $ee$ and $2.1\sigma$ for $\mu\mu$.
A dedicated experimental analysis of the HERA data and accurate theoretical estimation including full NLO corrections
needs be performed to understand the origin of these discrepancies and validate 
the computation of the lepton and photon PFDs of the proton and/or the electron.


\section{Resonant electroweak gauge boson production}
\label{resonant_production}
We consider now resonant production of electroweak bosons at HERA.
For the lepton-initiated process of resonantly producing $Z$-boson, we can calculate the value of $x_{p/e^+}$ that we expect to provide the largest contribution to the resonant cross-section from the relation involving the energy of the incoming particles: $4 E_e E_{p/e^+} \approx m_Z^2$, with $E_{p/e^+}=x_{p/e^+} E_{p}$. We find that at HERA the largest contribution comes from  $x_{p/e^+}\approx 0.082$  corresponding to $E_{p/e^+}\approx 75$ GeV. Therefore, considering the energy of the incoming electron, we expect the peak of the total energy distribution of the $Z$-boson decaying products to be around 103~GeV.

\begin{figure}[t]
\begin{center}
\includegraphics[ height=0.5\textwidth, width=0.7\linewidth]{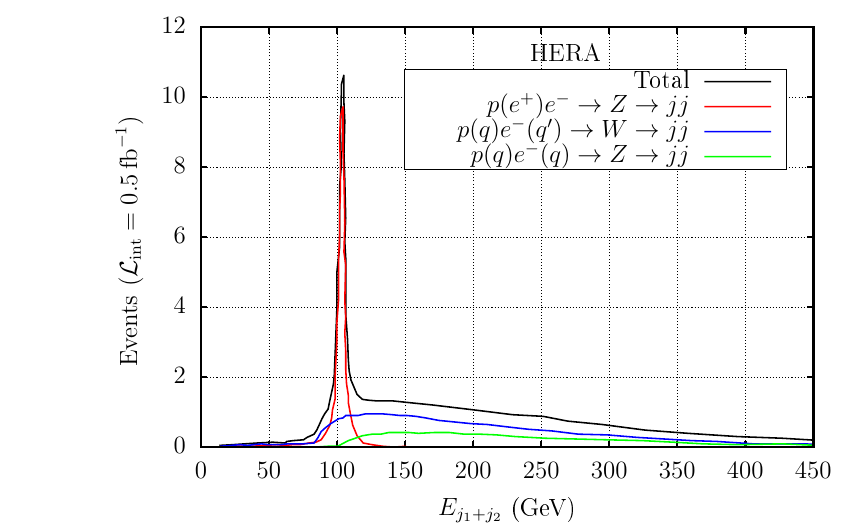}
\caption{Distribution of the sum of energies of the leading two jets, $E_{j_1 + j_2}$ (GeV), coming from $Z$- and $W^{\pm}$-bosons at HERA.} \label{Ej1j2}
\end{center}
\end{figure}
\begin{figure}[t!]
\begin{center}
\includegraphics[width=0.6\linewidth]{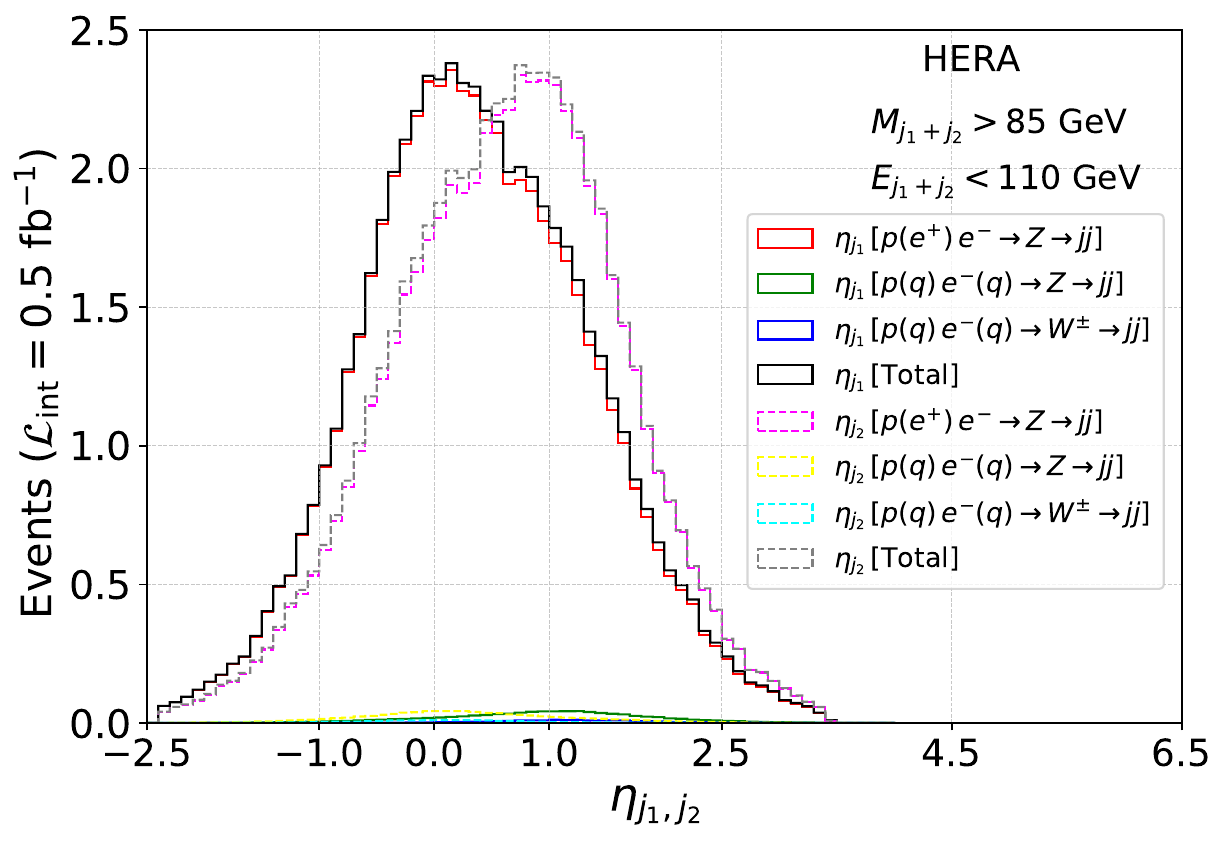}
\caption{$\eta$ distributions of the leading two jets ($j_1$ and $j_2$) originating from  the $Z$ and $W$-bosons at HERA after applying the selection cuts of $P_{T_{j_1, j_2}} > 5$~GeV, $M_{j_1 j_2} > 85$~GeV and   $E_{j_1 + j_2} < \, 110$~GeV.} \label{eta_dis}
\end{center}
\end{figure}
Observe that, contrary to the lepton-initiated process in which the main contribution comes from $x_{e^{-}/e^{-}}\simeq 1$ for the colliding electron, for the quark-initiated processes the longitudinal momenta of the quarks within the proton are further spread and thus 
it is expected that a range of  $x_{e^{-}/q}$ and  $x_{p/q}$ exists for which significant contributions to the resonant productions of $Z$ and also $W^{^\pm}$ may occur. We observe that quark-initiated production processes predominantly take place for $0.1\lesssim x_{e^{-}/q}\lesssim 0.5$.

In the case of the $Z$-resonance production, in order to maximize the signal, we focus on dijet final states for the $Z$-decay products. In Fig.~\ref{Ej1j2}, we present the total energy distribution  $E_{j_1 + j_2}$ of jets in resonant $Z$ and $W^\pm$ production, where $j_1$ and $j_2$ proceed from the decay of the produced $Z$ and $W^{\pm}$-bosons via the lepton-initiated (red) and the quark-initiated (green for $Z$ and blue for $W^{\pm}$) processes. One can clearly see the behaviour predicted from simple arguments in the previous paragraph. The spread of the quark-initiated process towards $E_{j_1 + j_2} > 103.5$~GeV is expected since larger values of $x_{p/q}$ in the proton provide larger contributions in the convolution than the mostly small and fixed value of $x_{p/e^+}\sim 0.082$,  implying a larger momentum for the jets coming from the decaying gauge boson. 

Note that  cuts on the distributions of $E_{j_1 + j_2}$ and the  invariant mass of the di-jets ($M_{j_1 j_2}$) can be applied to separate out the di-jet events coming from the
lepton-initiated $Z$ production process from the quark-initiated $Z$ and $W$ production processes.
In Fig.~\ref{eta_dis}, we present the $\eta$ distributions of the two leading jets ($\eta_{j_1}$ and $\eta_{j_2}$),
after applying the kinematical selection cuts of $M_{j_1 j_2} > 85$~GeV and $E_{j_1 + j_2} < 110$~GeV. These cuts prove highly effective, with almost negligible contributions from the quark-initiated processes, as can be seen via the distributions from the sum contribution (black and grey lines), which mostly overlap with the lepton-initiated process. It can also be seen that these jets are pretty central.

\begin{figure}[t!]
\begin{center}
\includegraphics[width=0.6\linewidth]{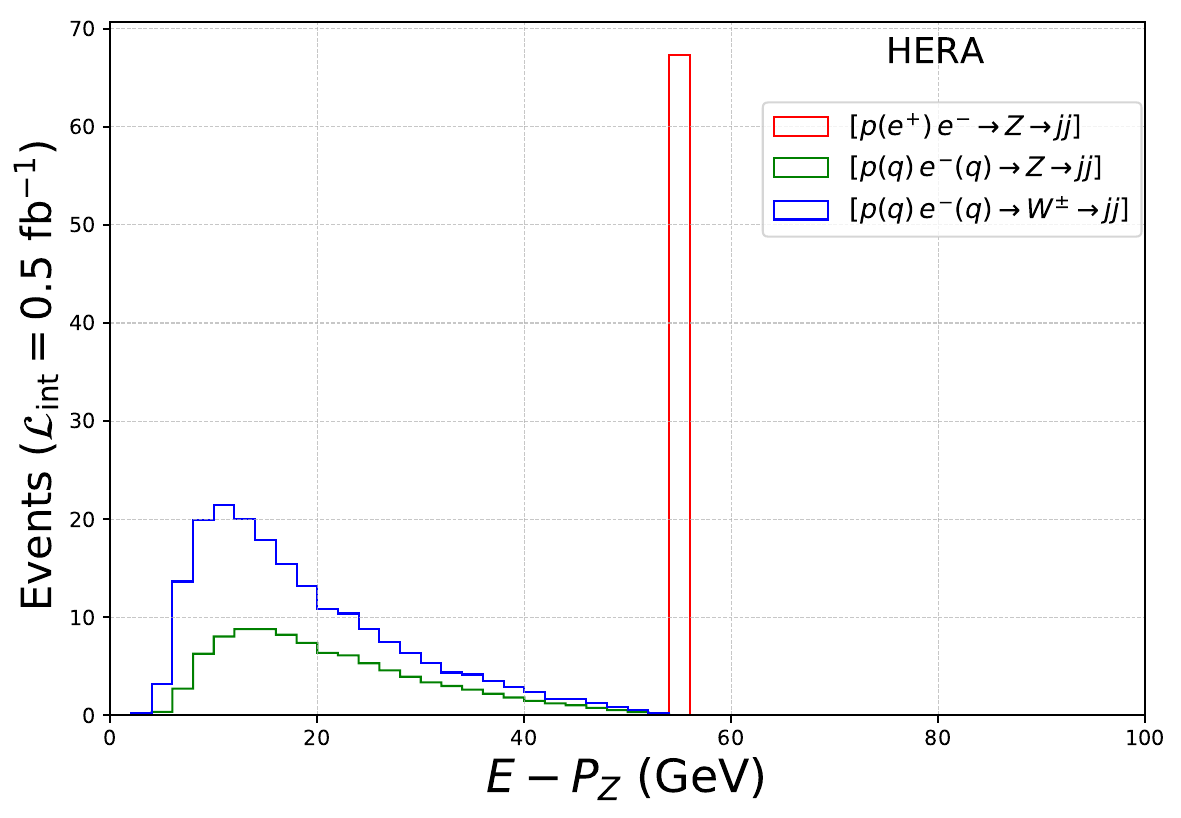}
\caption{Distribution of $E-P_z$ of decaying objects (in this case two-jets) of $Z$, $W^{\pm}$ bosons. Individual LO production processes, i.e., lepton-initiated and quark-initiated, of these gauge bosons are denoted by different colors.} \label{E_PZ}
\end{center}
\end{figure}

In addition to these cuts,  the  $E-P_z$ kinematical variable can be used to distinguish the lepton-initiated process from the quark-initiated processes.
We show the distribution $E-P_z$ in Fig.~\ref{E_PZ} for lepton- and quark-initiated production processes for resonant $Z$ and $W$ boson production.
Whereas for initial leptons a very localized peak around 2$E_e$ is obtained, quark-initiated events are spread and the maximum shows at $E-P_z \approx 10$ GeV.
Hence, implementing $ E - P_z > 52$~GeV cut can effectively serve to reject quark-initiated processes. Furthermore, as we will see later on,  this cut aids in the rejection of non-resonant dijet production which involves the photon PDF of electron~\cite{ZEUS:2008hcd,H1:2009rfg}.

Since we are interested in the di-jet final state from the resonant production of $Z$, we demand at least two jets with the characteristics described above in order to be associated with a $Z$-boson decay and we veto any leptons in the final state. Note however that in a $pe$ collider, a di-jet background with no lepton can arise from processes such as $p e^{-} \rightarrow j e^{-}$, $j j e^{-}$ and $j e^{-} \gamma$, where the final state electron is misidentified as a jet.  Focusing on events in the central region, i.e., $|\eta| < 2$, and considering the electron to be hard ($P_T > 20$ GeV), the fake rate of an electron being identified as a jet is of the order of $\lesssim 10^{-3}$ at HERA~\cite{Wessels:2004yz,communication}. We conducted an analysis considering this fake rate ($10^{-3}$) to estimate the number of background events from this source in our signal region and found it to be negligible. 

\begin{figure}[t!]
\begin{center}
\includegraphics[width=0.85\linewidth]{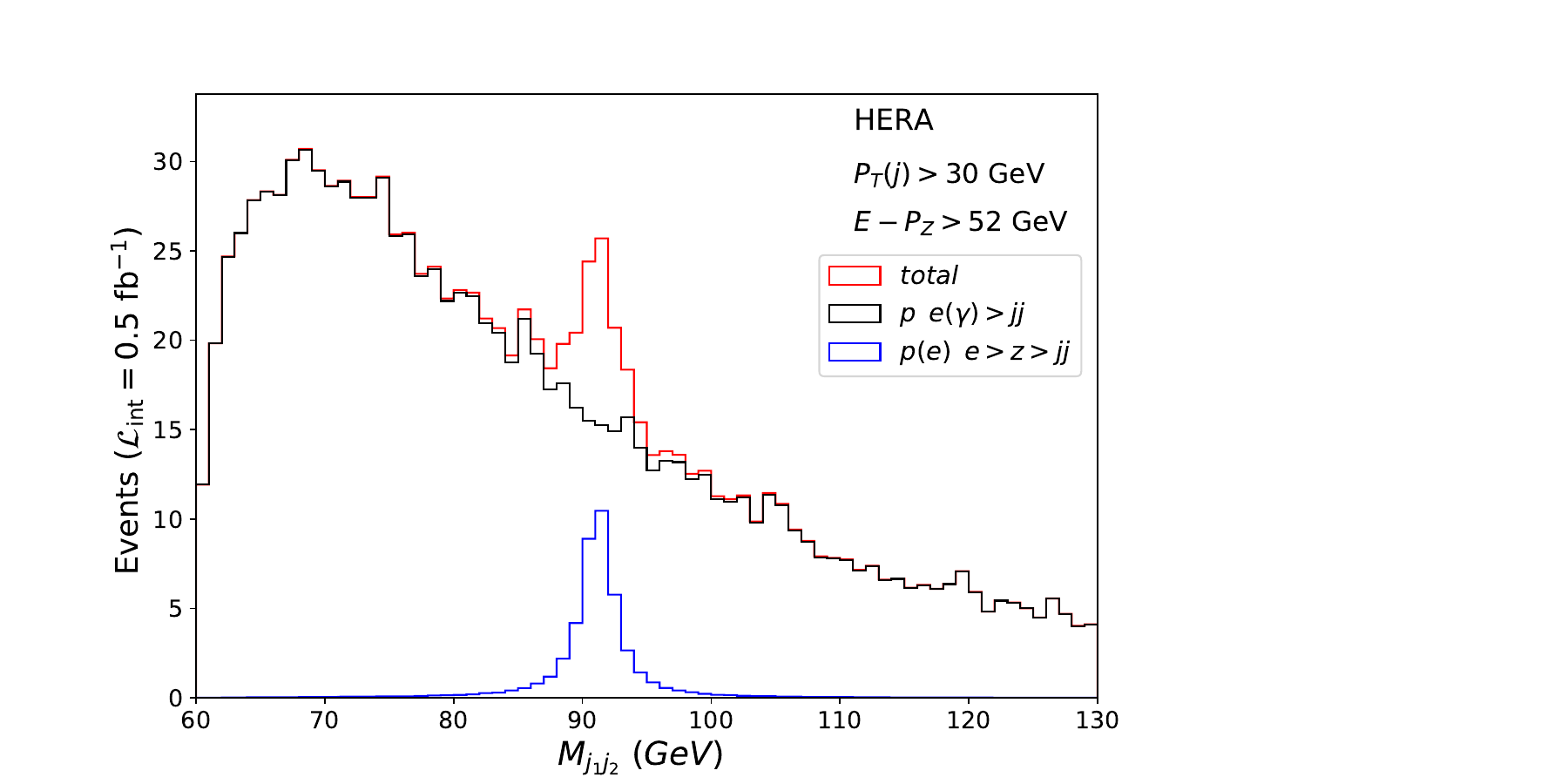}
\caption{
$M_{j_1 j_2}$ distribution originating from $pe^-(\gamma)\to jj$ (black) and $p(e^+)e^-\to Z\to jj$ (blue) processes at HERA under the selection cuts of $P_T^j  > 30$~GeV and $E-P_z > 52$~GeV. The total distributions are shown in red.} \label{jj_largex}
\end{center}
\end{figure}

One must consider other production processes of di-jet events at the $pe$ colliders that enter as background to our signal. Dominant contributions are generated involving the $\gamma$ and $q$ coming from the colliding electron and the $g$, $q$ and $\gamma$ coming from the colliding proton. Indeed, rejecting events for $E -P_z < 52$~GeV at HERA, selects an incoming parton from the electron with energy larger than 26 GeV, which leads to $x_{e^-/\gamma},\;  x_{e^-/q} > 0.942$. Since the photon PDF is not suppressed near $x_{e/\gamma} \simeq 1$ (see the purple line in Fig.~\ref{proton-elec-pdf}), a significant number of di-jet events can be generated in the same region of $E-P_z$ where the $Z$-boson is produced resonantly. In Fig.~\ref{jj_largex}, we present the distributions of
$M_{j_1 j_2}$ coming from the processes $pe^-(\gamma) \to j j$ 
and $p(e^+)e^- \to Z \to j j$ at HERA, under the selection cuts: $P_T^j  > 30$~GeV and $E-P_z > 52$~GeV. 

Considering di-jet events coming from the lepton-initiated Z-boson production as signal and other sources as background, we perform a cut-based analysis applying the following selection cuts: $E-P_z > 52$~GeV, $P_j > 30$~GeV, $|\eta_j| <~2$ and a 5~GeV selection window of $M_{j_1 j_2}$ and $E_{j_1 + j_2}$ around 91~GeV and 103~GeV, respectively.
We set $Q = m_Z / 2$ in our calculation. As discussed in section~\ref{NLO}, we vary $Q$ by a factor of 2 up and down, observing fluctuations of approximately 18\% for the signal. We obtain $55^{+9}_{-10}$ signal events,  and $162^{+13}_{-12}$ background events considering the combined integrated luminosity of 0.94~fb$^{-1}$ of H1 and ZEUS detectors~\cite{Nobe:2014csa}. This would correspond to a signal significance of around 4.3$\sigma$ at HERA, which should be easily identified in the HERA stored data.

Our estimation of the di-jet background at $E-P_z > 52$~GeV depends on QCD processes via, for example,  the non-resonant production of di-jets from the gluon  PDF of the $p$ and the photon PDF of the $e$. Since our calculations are at LO in QCD and QED and no proper shower, hadronization and detector effects have been taken into account, 
 it is better to obtain the di-jet background from data by extrapolating the distribution to the relevant region from the sidebands, in a similar approach as the one performed in the analysis of the di-photon background during the Higgs discovery at the Large Hadron Collider (LHC)~\cite{Jenni:2015hvm}.
This procedure would also allow to test the photon PDFs of the $e$ at $x_{e^-/\gamma} > 0.942$.

Analogously using sidebands to estimate the background, one could potentially study the $q$ PDFs of the $e$ by analyzing the di-jet events originating from the Z-boson and W-boson via the quark-initiated processes. As discussed, this production mechanism is enhanced in regions of smaller values of $E-P_z$ than the ones where lepton-initiated processes dominate, see Fig.~\ref{E_PZ}, implying at the same time a significantly larger di-jet background due to the growth of the photon PDF of the electron.

\section{EIC}
\label{EIC}
A similar analysis can be done with a future proposed experiment like the Electron-Ion collider (EIC). The EIC~\cite{Accardi:2012qut, AbdulKhalek:2021gbh}, to be built at Brookhaven National Laboratory, will collide polarized electrons and polarized protons/ions across a wide energy spectrum with high luminosities. Anticipated configurations involve a polarized electron beam with energies up to 21 GeV colliding with a polarized proton at 250 GeV, covering a center-of-mass (CoM) energy range from 30 GeV to 145 GeV with high luminosity. In this study, we simulate events with incoming proton and electron energies set at 250 GeV and 20 GeV, respectively and a luminosity of 100~$\text{fb}^{-1}$.

Similarly to our HERA analysis,
di-leptonic final states arising from non-resonant processes can also be studied at the EIC to test these PDFs.
 We estimate around $1.2 \times 10^4$, $1.9 \times 10^4$ and  $6 \times 10^2$ events at the LO in $e^{+} e^{-}$, $e^{-} e^{-}$ and $\mu^{+} \mu^{-}$ final states, respectively, at the EIC considering the selection cuts of $P_T^{\ell} > 10$~GeV and $|\eta_{\ell}| < 2.0$ and setting the renormalization scale as the scalar sum of $P_T/2$ of the leptons. Furthermore, considering the same cuts, one can expect $2.5 \times 10^4$  $e\mu$ events at the partonic level where approximately half of these events correspond to the same sign with different flavor. Note that, all these di-leptonic events correspond to $E-P_z = 2 E_e \sim 40$~GeV.
Additionally, same-flavor, opposite-charge di-lepton final states can arise from photo-production, predominantly in the regions of $0.1 \lesssim x_{e^{-}/\gamma} \lesssim 1.0$ and $0.04 \lesssim x_{p/\gamma} \lesssim 0.2$, under the same phase space cuts on $P_T^{\ell}$ and $\eta_{\ell}$, at the EIC. We estimate approximately $2.4\times 10^4$ $e^{+} e^{-}$ total events at the LO under the same cuts, out of which $2.7\times 10^3$ events occur in the $E-P_Z>35$~GeV region~\footnote{Similar number of events are expected for $\mu^{+} \mu^{-}$.}.
Once future data is available, a detailed study of these processes should be done for the EIC to test the lepton PDFs of the proton.
 
%
It is worth mentioning that these findings could be further tested by considering non-resonant di-lepton final states of different flavours and/or the same charge at the LHC~\cite{Buonocore:2020nai}.
Although the cross-sections of such processes are small, a significant number of events at the high luminosity (HL)-LHC are expected due to its higher luminosity. 
Such studies, using forthcoming data from the LHC, could probe the lepton PDFs of the proton.
 Moreover, a more precise estimation of the SM backgrounds derived from such analyses could be instrumental in probing beyond the SM scenarios that involve these specific final states at the LHC.

In addition to studying the non-resonant di-leptonic final states, one can also investigate resonant $Z$ boson production at the EIC. Focusing on the lepton-initiated process, and considering that the electron carries an energy of around 20 GeV, the $e^{+}$ parton within the proton needs an energy of approximately 103.5 GeV, corresponding to $x_{p/e^{+}} \sim 0.41$, to produce the Z-boson resonantly. This implies that the distribution of the energy sum of the leading two jets originating from the $Z$-boson decay, generated through the lepton-initiated process, is expected to peak around 123.5 GeV at the EIC.
The quark-initiated processes can also generate $Z$ and $W$-bosons resonantly at EIC. This corresponds to different x-regions which are $0.41 \lesssim x_{p/q}, x_{e^{-}/q}  \lesssim 1$ and $0.32 \lesssim x_{p/q}, x_{e^{-}/q}  \lesssim 1$ for the $Z$ and $W$-boson productions, respectively. Thus, the EIC exhibits sensitivity to relatively larger `x'-regions compared to what HERA could probe.

We estimate the signal significance for the lepton-initiated Z-boson production in the di-jet decay mode using the following selection cuts: $P_T^j >$ 30 GeV, $|\eta_j| < 2$, $E-P_z > 36$~GeV and a 5 GeV selection window of $M_{j_1 j_2}$ and  $E_{j_1 + j_2}$
around 91 GeV and 123 GeV, respectively, obtaining 232 signal events and 3210  background events coming from the photo-production. This corresponds to a signal significance of around 4.1$\sigma$. Since the di-jet background remains significantly larger than the signal even when focusing on the relevant region of invariant masses around $m_Z$, one should use the background subtraction method to deal with the background properly. A different set of cuts on $M_{j_1 j_2}$,  $E_{j_1 + j_2}$ and $E-P_z$ can be applied to study the quark-initiated processes, i.e., to study the entire range $0.32 \lesssim x_{p/q}, x_{e^{-}/q}  \lesssim 1$.

Due to the larger luminosity at the EIC compared with HERA, it is in principle possible the study of $Z$-boson resonant production in its leptonic decay mode, taking advantage of that the non-resonant di-lepton background is expected to be small, particularly, if both leptons are taken to be muons. We estimate that the quark-initiated (lepton-initiated) resonantly produced $Z$-boson can yield 11 (22) signal events each in the di-muon and di-electron final states with the selection cuts of $P_T^{\ell} >$ 10 GeV, $|\eta_{\ell}| < 2$.
 Backgrounds for the $\mu^{+} \mu^{-}$ final states are dominated by photo-production. 
Considering these cuts with a 5 GeV selection window for $M_{\mu^{+} \mu^{-}}$ and $E_{\mu^{+} \mu^{-}}$ around 91 GeV and 123 GeV, respectively, as well as the $E-P_z > 35$~GeV selection cuts, we obtain 2--3 events from the photo-production whereas 15 events remain from the resonant Z production through lepton-initiated process at the EIC. Thus, this could be a much cleaner channel to study lepton-initiated resonant $Z$ production compared with the final states involving di-jets and $e^{+} e^{-}$. A dedicated analysis should be performed once future data is available. 
%
\section{Conclusions}
\label{Conclusions}
In this article we introduce a novel approach to validate the lepton content of the proton by analyzing both non-resonant di-lepton production and di-jets from a resonantly produced $Z$ gauge boson, using existing HERA data and future measurements from the EIC. For $Z$-resonant production, we propose conducting this study at $pe$ colliders rather than at a hadron collider like the LHC, as the lepton-initiated process, which depends on the lepton PDFs of the proton, can be more easily separated from the quark-initiated ones. 
It is also feasible to study the leptonic decay modes of the $Z$ boson, particularly in the di-muon final state, at the EIC. This is due to the substantial number of expected signal events, thanks to the large integrated luminosity of the EIC and relatively low di-muon background. 

In the case of non-resonant processes involving di-lepton final states, we cover all possible combinations of the same and different flavors, as well as the same and opposite charges. These studies can explore a  relatively lower $x_{p/e,\mu}$ of the lepton PDFs of proton than the resonant $Z$-boson production case, providing valuable insights into the distribution of leptons within the proton. 
Comparisons with the published HERA experimental data suggest  overestimation of the production rates for the di-lepton final states by a few tens of percent, commensurate with uncertainties  in our LO computation.
A dedicated experimental analysis of HERA data, combined with precise theoretical estimations incorporating higher-order corrections, is essential to accurately identify the origin of these discrepancies and validate the PDFs computations.
Additionally, we suggest similar studies with future data from the EIC and HL-LHC, which could further test the estimation of the lepton PDFs of the proton.

Let us comment in closing that the idea proposed in this paper, studying resonant 
$Z$ production and non-resonant di-leptonic final states, can also be pursued in future experiments, such as the Muon-Ion Collider (MuIC)~\cite{Acosta:2021qpx}, Large Hadron electron Collider (LHeC)~\cite{LHeC:2020van} and Future Circular lepton-hadron collider (FCC-eh)~\cite{Bruning:2022hro}, if they are realized.  The much higher CoM energy of these facilities
 would allow for the exploration of different regions of the lepton PDFs of the proton. 
\section*{Acknowledgements} 
We thank Luca Buonocore, Sergei Chekanov, Daniel de Florian, Tao Han, Tim Hobbs, Ian Low, David Marzocca, LianTao Wang, Martin Wessels  and Rik Yoshida for usueful discussions. We also thank Paolo Nason for help with LUXlep. 
We extend special thanks to Tim Hobbs for his careful reading of the manuscript and his valuable comments.
The work of LD has been partially supported by CONICET and ANPCyT projects PIP-11220200101426
and PICT-2018-03682. The work of AM has been partially supported by CONICET and the ANPCyT project  PICT-2018-03682. SR and CW\ are supported by the U.S.~Department of Energy under contracts No.\ DEAC02- 06CH11357 at Argonne National Laboratory. 
The work of CW\ at the University of Chicago has been supported by the DOE grant DE-SC0013642.
AM thanks CW and the hospitality
from the EFI at the University of Chicago and the HEP division at ANL where part of this work
was done.
SR would like to thank the University of Chicago and Fermilab, where a significant part of this work has been done.  C.W.\ would like to thank the Aspen Center for Physics, which is supported by National Science Foundation grant No.~PHY-1607611, where part of this work has been done.
\bibliographystyle{utphys}
\bibliography{reference.bib}

\providecommand{\href}[2]{#2}\begingroup\raggedright\begin{thebibliography}{10}

\bibitem{Bjorken:1968dy}
J.~D. Bjorken, ``{Asymptotic Sum Rules at Infinite Momentum},'' \href{http://dx.doi.org/10.1103/PhysRev.179.1547}{{\em Phys. Rev.} {\bfseries 179} (1969) 1547--1553}.

\bibitem{Feynman:1969wa}
R.~P. Feynman, ``{The behavior of hadron collisions at extreme energies},'' {\em Conf. Proc. C} {\bfseries 690905} (1969) 237--258.

\bibitem{Bjorken:1969ja}
J.~D. Bjorken and E.~A. Paschos, ``{Inelastic Electron Proton and gamma Proton Scattering, and the Structure of the Nucleon},'' \href{http://dx.doi.org/10.1103/PhysRev.185.1975}{{\em Phys. Rev.} {\bfseries 185} (1969) 1975--1982}.

\bibitem{Altarelli:1977zs}
G.~Altarelli and G.~Parisi, ``{Asymptotic Freedom in Parton Language},'' \href{http://dx.doi.org/10.1016/0550-3213(77)90384-4}{{\em Nucl. Phys. B} {\bfseries 126} (1977) 298--318}.

\bibitem{vonWeizsacker:1934nji}
C.~F. von Weizsacker, ``{Radiation emitted in collisions of very fast electrons},'' \href{http://dx.doi.org/10.1007/BF01333110}{{\em Z. Phys.} {\bfseries 88} (1934) 612--625}.

\bibitem{Williams:1934ad}
E.~J. Williams, ``{Nature of the high-energy particles of penetrating radiation and status of ionization and radiation formulae},'' \href{http://dx.doi.org/10.1103/PhysRev.45.729}{{\em Phys. Rev.} {\bfseries 45} (1934) 729--730}.

\bibitem{Gribov:1972ri}
V.~N. Gribov and L.~N. Lipatov, ``{Deep inelastic e p scattering in perturbation theory},'' {\em Sov. J. Nucl. Phys.} {\bfseries 15} (1972) 438--450.

\bibitem{Dokshitzer:1977sg}
Y.~L. Dokshitzer, ``{Calculation of the Structure Functions for Deep Inelastic Scattering and e+ e- Annihilation by Perturbation Theory in Quantum Chromodynamics.},'' {\em Sov. Phys. JETP} {\bfseries 46} (1977) 641--653.

\bibitem{Garosi:2023bvq}
F.~Garosi, D.~Marzocca, and S.~Trifinopoulos, ``{LePDF: Standard Model PDFs for high-energy lepton colliders},'' \href{http://dx.doi.org/10.1007/JHEP09(2023)107}{{\em JHEP} {\bfseries 09} (2023) 107}, \href{http://arxiv.org/abs/2303.16964}{{\ttfamily arXiv:2303.16964 [hep-ph]}}.

\bibitem{Han:2020uid}
T.~Han, Y.~Ma, and K.~Xie, ``{High energy leptonic collisions and electroweak parton distribution functions},'' \href{http://dx.doi.org/10.1103/PhysRevD.103.L031301}{{\em Phys. Rev. D} {\bfseries 103} no.~3, (2021) L031301}, \href{http://arxiv.org/abs/2007.14300}{{\ttfamily arXiv:2007.14300 [hep-ph]}}.

\bibitem{Han:2021kes}
T.~Han, Y.~Ma, and K.~Xie, ``{Quark and gluon contents of a lepton at high energies},'' \href{http://dx.doi.org/10.1007/JHEP02(2022)154}{{\em JHEP} {\bfseries 02} (2022) 154}, \href{http://arxiv.org/abs/2103.09844}{{\ttfamily arXiv:2103.09844 [hep-ph]}}.

\bibitem{Greco:2016izi}
M.~Greco, T.~Han, and Z.~Liu, ``{ISR effects for resonant Higgs production at future lepton colliders},'' \href{http://dx.doi.org/10.1016/j.physletb.2016.10.078}{{\em Phys. Lett. B} {\bfseries 763} (2016) 409--415}, \href{http://arxiv.org/abs/1607.03210}{{\ttfamily arXiv:1607.03210 [hep-ph]}}.

\bibitem{Spiesberger:1994dm}
H.~Spiesberger, ``{QED radiative corrections for parton distributions},'' \href{http://dx.doi.org/10.1103/PhysRevD.52.4936}{{\em Phys. Rev. D} {\bfseries 52} (1995) 4936--4940}, \href{http://arxiv.org/abs/hep-ph/9412286}{{\ttfamily arXiv:hep-ph/9412286}}.

\bibitem{Roth:2004ti}
M.~Roth and S.~Weinzierl, ``{QED corrections to the evolution of parton distributions},'' \href{http://dx.doi.org/10.1016/j.physletb.2004.04.009}{{\em Phys. Lett. B} {\bfseries 590} (2004) 190--198}, \href{http://arxiv.org/abs/hep-ph/0403200}{{\ttfamily arXiv:hep-ph/0403200}}.

\bibitem{Martin:2004dh}
A.~D. Martin, R.~G. Roberts, W.~J. Stirling, and R.~S. Thorne, ``{Parton distributions incorporating QED contributions},'' \href{http://dx.doi.org/10.1140/epjc/s2004-02088-7}{{\em Eur. Phys. J. C} {\bfseries 39} (2005) 155--161}, \href{http://arxiv.org/abs/hep-ph/0411040}{{\ttfamily arXiv:hep-ph/0411040}}.

\bibitem{Buonocore:2020nai}
L.~Buonocore, P.~Nason, F.~Tramontano, and G.~Zanderighi, ``{Leptons in the proton},'' \href{http://dx.doi.org/10.1007/JHEP08(2020)019}{{\em JHEP} {\bfseries 08} no.~08, (2020) 019}, \href{http://arxiv.org/abs/2005.06477}{{\ttfamily arXiv:2005.06477 [hep-ph]}}.

\bibitem{Manohar:2016nzj}
A.~Manohar, P.~Nason, G.~P. Salam, and G.~Zanderighi, ``{How bright is the proton? A precise determination of the photon parton distribution function},'' \href{http://dx.doi.org/10.1103/PhysRevLett.117.242002}{{\em Phys. Rev. Lett.} {\bfseries 117} no.~24, (2016) 242002}, \href{http://arxiv.org/abs/1607.04266}{{\ttfamily arXiv:1607.04266 [hep-ph]}}.

\bibitem{Manohar:2017eqh}
A.~V. Manohar, P.~Nason, G.~P. Salam, and G.~Zanderighi, ``{The Photon Content of the Proton},'' \href{http://dx.doi.org/10.1007/JHEP12(2017)046}{{\em JHEP} {\bfseries 12} (2017) 046}, \href{http://arxiv.org/abs/1708.01256}{{\ttfamily arXiv:1708.01256 [hep-ph]}}.

\bibitem{Ball:2013hta}
{\bfseries NNPDF} Collaboration, R.~D. Ball, V.~Bertone, S.~Carrazza, L.~Del~Debbio, S.~Forte, A.~Guffanti, N.~P. Hartland, and J.~Rojo, ``{Parton distributions with QED corrections},'' \href{http://dx.doi.org/10.1016/j.nuclphysb.2013.10.010}{{\em Nucl. Phys. B} {\bfseries 877} (2013) 290--320}, \href{http://arxiv.org/abs/1308.0598}{{\ttfamily arXiv:1308.0598 [hep-ph]}}.

\bibitem{Bertone:2017bme}
{\bfseries NNPDF} Collaboration, V.~Bertone, S.~Carrazza, N.~P. Hartland, and J.~Rojo, ``{Illuminating the photon content of the proton within a global PDF analysis},'' \href{http://dx.doi.org/10.21468/SciPostPhys.5.1.008}{{\em SciPost Phys.} {\bfseries 5} no.~1, (2018) 008}, \href{http://arxiv.org/abs/1712.07053}{{\ttfamily arXiv:1712.07053 [hep-ph]}}.

\bibitem{Cridge:2021pxm}
T.~Cridge, L.~A. Harland-Lang, A.~D. Martin, and R.~S. Thorne, ``{QED parton distribution functions in the MSHT20 fit},'' \href{http://dx.doi.org/10.1140/epjc/s10052-022-10028-2}{{\em Eur. Phys. J. C} {\bfseries 82} no.~1, (2022) 90}, \href{http://arxiv.org/abs/2111.05357}{{\ttfamily arXiv:2111.05357 [hep-ph]}}.

\bibitem{Xie:2021equ}
{\bfseries CTEQ-TEA} Collaboration, K.~Xie, T.~J. Hobbs, T.-J. Hou, C.~Schmidt, M.~Yan, and C.~P. Yuan, ``{Photon PDF within the CT18 global analysis},'' \href{http://dx.doi.org/10.1103/PhysRevD.105.054006}{{\em Phys. Rev. D} {\bfseries 105} no.~5, (2022) 054006}, \href{http://arxiv.org/abs/2106.10299}{{\ttfamily arXiv:2106.10299 [hep-ph]}}.

\bibitem{Xie:2023qbn}
{\bfseries CTEQ-TEA} Collaboration, K.~Xie, B.~Zhou, and T.~J. Hobbs, ``{The photon content of the neutron},'' \href{http://dx.doi.org/10.1007/JHEP04(2024)022}{{\em JHEP} {\bfseries 04} (2024) 022}, \href{http://arxiv.org/abs/2305.10497}{{\ttfamily arXiv:2305.10497 [hep-ph]}}.

\bibitem{NNPDF:2024djq}
{\bfseries NNPDF} Collaboration, R.~D. Ball {\em et~al.}, ``{Photons in the proton: implications for the LHC},'' \href{http://dx.doi.org/10.1140/epjc/s10052-024-12731-8}{{\em Eur. Phys. J. C} {\bfseries 84} no.~5, (2024) 540}, \href{http://arxiv.org/abs/2401.08749}{{\ttfamily arXiv:2401.08749 [hep-ph]}}.

\bibitem{ZEUS:2008goe}
{\bfseries ZEUS} Collaboration, S.~Chekanov {\em et~al.}, ``{Search for events with an isolated lepton and missing transverse momentum and a measurement of W production at HERA},'' \href{http://dx.doi.org/10.1016/j.physletb.2009.01.014}{{\em Phys. Lett. B} {\bfseries 672} (2009) 106--115}, \href{http://arxiv.org/abs/0807.0589}{{\ttfamily arXiv:0807.0589 [hep-ex]}}.

\bibitem{ZEUS:2003vfj}
{\bfseries ZEUS} Collaboration, S.~Chekanov {\em et~al.}, ``{Search for single top production in ep collisions at HERA},'' \href{http://dx.doi.org/10.1016/S0370-2693(03)00333-2}{{\em Phys. Lett. B} {\bfseries 559} (2003) 153--170}, \href{http://arxiv.org/abs/hep-ex/0302010}{{\ttfamily arXiv:hep-ex/0302010}}.

\bibitem{Kerger:2000eh}
{\bfseries ZEUS, H1} Collaboration, R.~Kerger, ``{Lepton flavor violation at HERA},'' in {\em {8th International Workshop on Deep Inelastic Scattering and QCD (DIS 2000)}}, pp.~426--429.
\newblock 6, 2000.
\newblock \href{http://arxiv.org/abs/hep-ex/0006023}{{\ttfamily arXiv:hep-ex/0006023}}.

\bibitem{H1:2006wzo}
{\bfseries H1} Collaboration, A.~Aktas {\em et~al.}, ``{Tau lepton production in ep collisions at HERA},'' \href{http://dx.doi.org/10.1140/epjc/s10052-006-0028-2}{{\em Eur. Phys. J. C} {\bfseries 48} (2006) 699--714}, \href{http://arxiv.org/abs/hep-ex/0604022}{{\ttfamily arXiv:hep-ex/0604022}}.

\bibitem{ZEUS:1999qvb}
{\bfseries ZEUS} Collaboration, J.~Breitweg {\em et~al.}, ``{W production and the search for events with an isolated high-energy lepton and missing transverse momentum at HERA},'' \href{http://dx.doi.org/10.1016/S0370-2693(99)01358-1}{{\em Phys. Lett. B} {\bfseries 471} (2000) 411--428}, \href{http://arxiv.org/abs/hep-ex/9907023}{{\ttfamily arXiv:hep-ex/9907023}}.

\bibitem{ZEUS:2003nmf}
{\bfseries ZEUS} Collaboration, S.~Chekanov {\em et~al.}, ``{Isolated tau leptons in events with large missing transverse momentum at HERA},'' \href{http://dx.doi.org/10.1016/j.physletb.2003.12.054}{{\em Phys. Lett. B} {\bfseries 583} (2004) 41--58}, \href{http://arxiv.org/abs/hep-ex/0311028}{{\ttfamily arXiv:hep-ex/0311028}}.

\bibitem{Diener:2002if}
K.-P.~O. Diener, C.~Schwanenberger, and M.~Spira, ``{Photoproduction of W bosons at HERA: QCD corrections},'' \href{http://dx.doi.org/10.1007/s10052-002-1023-x}{{\em Eur. Phys. J. C} {\bfseries 25} (2002) 405--411}, \href{http://arxiv.org/abs/hep-ph/0203269}{{\ttfamily arXiv:hep-ph/0203269}}.

\bibitem{ZEUS:1997bxs}
{\bfseries ZEUS} Collaboration, J.~Breitweg {\em et~al.}, ``{Comparison of ZEUS data with standard model predictions for $e^{+} p \to e^{+} X$ scattering at high x and $Q^{2}$},'' \href{http://dx.doi.org/10.1007/s002880050384}{{\em Z. Phys. C} {\bfseries 74} (1997) 207--220}, \href{http://arxiv.org/abs/hep-ex/9702015}{{\ttfamily arXiv:hep-ex/9702015}}.

\bibitem{ZEUS:2002pdo}
{\bfseries ZEUS} Collaboration, S.~Chekanov {\em et~al.}, ``{Measurement of high Q**2 e- p neutral current cross-sections at HERA and the extraction of xF(3)},'' \href{http://dx.doi.org/10.1140/epjc/s2003-01163-y}{{\em Eur. Phys. J. C} {\bfseries 28} (2003) 175--201}, \href{http://arxiv.org/abs/hep-ex/0208040}{{\ttfamily arXiv:hep-ex/0208040}}.

\bibitem{ZEUS:2008hcd}
{\bfseries ZEUS} Collaboration, S.~Chekanov {\em et~al.}, ``{A Measurement of the Q**2, W and t dependences of deeply virtual Compton scattering at HERA},'' \href{http://dx.doi.org/10.1088/1126-6708/2009/05/108}{{\em JHEP} {\bfseries 05} (2009) 108}, \href{http://arxiv.org/abs/0812.2517}{{\ttfamily arXiv:0812.2517 [hep-ex]}}.

\bibitem{H1:2009rfg}
{\bfseries H1} Collaboration, F.~D. Aaron {\em et~al.}, ``{Events with Isolated Leptons and Missing Transverse Momentum and Measurement of $W$ Production at HERA},'' \href{http://dx.doi.org/10.1140/epjc/s10052-009-1160-6}{{\em Eur. Phys. J. C} {\bfseries 64} (2009) 251--271}, \href{http://arxiv.org/abs/0901.0488}{{\ttfamily arXiv:0901.0488 [hep-ex]}}.

\bibitem{H1:2009mij}
{\bfseries H1, ZEUS} Collaboration, F.~D. Aaron {\em et~al.}, ``{Multi-Leptons with High Transverse Momentum at HERA},'' \href{http://dx.doi.org/10.1088/1126-6708/2009/10/013}{{\em JHEP} {\bfseries 10} (2009) 013}, \href{http://arxiv.org/abs/0907.3627}{{\ttfamily arXiv:0907.3627 [hep-ex]}}.

\bibitem{Nobe:2014csa}
{\bfseries H1, ZEUS} Collaboration, T.~Nobe, ``{Real $W$ and $Z$ bosons production at HERA},'' \href{http://dx.doi.org/10.1051/epjconf/20147100099}{{\em EPJ Web Conf.} {\bfseries 71} (2014) 00099}.

\bibitem{Clark:2016jgm}
D.~B. Clark, E.~Godat, and F.~I. Olness, ``{ManeParse : A Mathematica reader for Parton Distribution Functions},'' \href{http://dx.doi.org/10.1016/j.cpc.2017.03.004}{{\em Comput. Phys. Commun.} {\bfseries 216} (2017) 126--137}, \href{http://arxiv.org/abs/1605.08012}{{\ttfamily arXiv:1605.08012 [hep-ph]}}.

\bibitem{Buonocore:2021bsf}
L.~Buonocore, P.~Nason, F.~Tramontano, and G.~Zanderighi, ``{Photon and leptons induced processes at the LHC},'' \href{http://dx.doi.org/10.1007/JHEP12(2021)073}{{\em JHEP} {\bfseries 12} (2021) 073}, \href{http://arxiv.org/abs/2109.10924}{{\ttfamily arXiv:2109.10924 [hep-ph]}}.

\bibitem{Sjostrand:2014zea}
T.~Sj\"ostrand, S.~Ask, J.~R. Christiansen, R.~Corke, N.~Desai, P.~Ilten, S.~Mrenna, S.~Prestel, C.~O. Rasmussen, and P.~Z. Skands, ``{An introduction to PYTHIA 8.2},'' \href{http://dx.doi.org/10.1016/j.cpc.2015.01.024}{{\em Comput. Phys. Commun.} {\bfseries 191} (2015) 159--177}, \href{http://arxiv.org/abs/1410.3012}{{\ttfamily arXiv:1410.3012 [hep-ph]}}.

\bibitem{Buonocore:2022msy}
L.~Buonocore, A.~Greljo, P.~Krack, P.~Nason, N.~Selimovic, F.~Tramontano, and G.~Zanderighi, ``{Resonant leptoquark at NLO with POWHEG},'' \href{http://dx.doi.org/10.1007/JHEP11(2022)129}{{\em JHEP} {\bfseries 11} (2022) 129}, \href{http://arxiv.org/abs/2209.02599}{{\ttfamily arXiv:2209.02599 [hep-ph]}}.

\bibitem{Almeida:2022udp}
E.~d.~S. Almeida, A.~Alves, O.~J.~P. \'Eboli, and F.~S. Queiroz, ``{Resonant lepton-gluon collisions at the Large Hadron Collider},'' \href{http://dx.doi.org/10.1103/PhysRevD.107.055024}{{\em Phys. Rev. D} {\bfseries 107} no.~5, (2023) 055024}, \href{http://arxiv.org/abs/2212.06178}{{\ttfamily arXiv:2212.06178 [hep-ph]}}.

\bibitem{Korajac:2023xtv}
A.~Korajac, P.~Krack, and N.~Selimovic, ``{Third-family lepton-quark fusion},'' \href{http://dx.doi.org/10.1140/epjc/s10052-024-12618-8}{{\em Eur. Phys. J. C} {\bfseries 84} no.~3, (2024) 304}, \href{http://arxiv.org/abs/2311.13635}{{\ttfamily arXiv:2311.13635 [hep-ph]}}.

\bibitem{Alwall:2014hca}
J.~Alwall, R.~Frederix, S.~Frixione, V.~Hirschi, F.~Maltoni, O.~Mattelaer, H.~S. Shao, T.~Stelzer, P.~Torrielli, and M.~Zaro, ``{The automated computation of tree-level and next-to-leading order differential cross sections, and their matching to parton shower simulations},'' \href{http://dx.doi.org/10.1007/JHEP07(2014)079}{{\em JHEP} {\bfseries 07} (2014) 079}, \href{http://arxiv.org/abs/1405.0301}{{\ttfamily arXiv:1405.0301 [hep-ph]}}.

\bibitem{Araz:2020lnp}
J.~Y. Araz, B.~Fuks, and G.~Polykratis, ``{Simplified fast detector simulation in MADANALYSIS 5},'' \href{http://dx.doi.org/10.1140/epjc/s10052-021-09052-5}{{\em Eur. Phys. J. C} {\bfseries 81} no.~4, (2021) 329}, \href{http://arxiv.org/abs/2006.09387}{{\ttfamily arXiv:2006.09387 [hep-ph]}}.

\bibitem{Wessels:2004yz}
M.~Wessels, \href{http://dx.doi.org/10.3204/DESY-THESIS-2004-035}{{\em {General Search for New Phenomena in $ep$ Scattering at HERA}}}.
\newblock PhD thesis, Aachen, Tech. Hochsch., 2004.

\bibitem{communication}
``{Private communication with Martin Wessels and Rik Yoshida},''.

\bibitem{Jenni:2015hvm}
P.~Jenni and T.~S. Virdee, ``{The Discovery of the Higgs Boson at the LHC},'' \href{http://dx.doi.org/10.1142/9789814644150_0001}{{\em Adv. Ser. Direct. High Energy Phys.} {\bfseries 23} (2015) 1--30}.

\bibitem{Accardi:2012qut}
A.~Accardi {\em et~al.}, ``{Electron Ion Collider: The Next QCD Frontier}: {Understanding the glue that binds us all},'' \href{http://dx.doi.org/10.1140/epja/i2016-16268-9}{{\em Eur. Phys. J. A} {\bfseries 52} no.~9, (2016) 268}, \href{http://arxiv.org/abs/1212.1701}{{\ttfamily arXiv:1212.1701 [nucl-ex]}}.

\bibitem{AbdulKhalek:2021gbh}
R.~Abdul~Khalek {\em et~al.}, ``{Science Requirements and Detector Concepts for the Electron-Ion Collider}: {EIC Yellow Report},'' \href{http://dx.doi.org/10.1016/j.nuclphysa.2022.122447}{{\em Nucl. Phys. A} {\bfseries 1026} (2022) 122447}, \href{http://arxiv.org/abs/2103.05419}{{\ttfamily arXiv:2103.05419 [physics.ins-det]}}.

\bibitem{Acosta:2021qpx}
D.~Acosta and W.~Li, ``{A muon\textendash{}ion collider at BNL: The future QCD frontier and path to a new energy frontier of \ensuremath{\mu}+\ensuremath{\mu}\ensuremath{-} colliders},'' \href{http://dx.doi.org/10.1016/j.nima.2022.166334}{{\em Nucl. Instrum. Meth. A} {\bfseries 1027} (2022) 166334}, \href{http://arxiv.org/abs/2107.02073}{{\ttfamily arXiv:2107.02073 [physics.acc-ph]}}.

\bibitem{LHeC:2020van}
{\bfseries LHeC, FCC-he Study Group} Collaboration, P.~Agostini {\em et~al.}, ``{The Large Hadron\textendash{}Electron Collider at the HL-LHC},'' \href{http://dx.doi.org/10.1088/1361-6471/abf3ba}{{\em J. Phys. G} {\bfseries 48} no.~11, (2021) 110501}, \href{http://arxiv.org/abs/2007.14491}{{\ttfamily arXiv:2007.14491 [hep-ex]}}.

\bibitem{Bruning:2022hro}
O.~Br\"uning, A.~Seryi, and S.~Verd\'u-Andr\'es, ``{Electron-Hadron Colliders: EIC, LHeC and FCC-eh},'' \href{http://dx.doi.org/10.3389/fphy.2022.886473}{{\em Front. in Phys.} {\bfseries 10} (2022) 886473}.

\end{thebibliography}\endgroup
\end{document}